\documentclass[hyper,12pt,a4paper]{article}
\usepackage{mathrsfs, subfigure}
\usepackage{amsmath}
\usepackage{amsfonts}
\usepackage{amssymb}
\usepackage{graphicx}

\usepackage{jheppub}

\newcommand{\be}{\begin{equation}}
\newcommand{\bea}{\begin{eqnarray}}
\newcommand{\eea}{\end{eqnarray}}
\newcommand{\ba}{\begin{array}}
\newcommand{\ea}{\end{array}}
\newcommand{\ee}{\end{equation}}

\def\bse{\begin{subequations}}
\def\ese{\end{subequations}}

\title{Holographic model of hybrid and coexisting s-wave and p-wave Josephson junction}

\author{Shuai Liu,}
\author{Yong-Qiang Wang}

\affiliation{Institute of Theoretical Physics, Lanzhou University,
Lanzhou 730000, People's Republic of China}
\emailAdd{liush13@lzu.edu.cn,yqwang@lzu.edu.cn}

\abstract{
In this paper the holographic model for a hybrid and coexisting s-wave
and p-wave Josephson junction is constructed by a triplet charged scalar field coupled
with a non-Abelian $SU(2)$ gauge field in (\(3+1\))-dimensional AdS spacetime.
Depending on the value of chemical potential $\mu$, one can show that there are four
types of junctions (s+p-N-s+p, s+p-N-s, s+p-N-p and s-N-p). We show that the DC currents
of all the hybrid and coexisting s-wave and p-wave junctions are proportional
to the sine of the phase difference across the junction. In addition, the maximum
current and the total condensation decay with the width of junction exponentially,
respectively. For the s+p-N-s and s-N-p junctions, the maximum current decreases with
growing temperature. Moreover, we find that the maximum current increases with
growing temperature for the s+p-N-s+p and s+p-N-p junctions, which is  different
from the behavior of the s+p-N-s and s-N-p junctions.}

\keywords{}
\begin{document}
\maketitle

\section{Introduction}

The study of superconductivity has been at the forefront of condensed matter physics.
In particular, what the origin is of high temperature superconductivity is still one
of the major unsolved problems of condensed matter theory. Over the past decade,
one of the most important result in string theory is the AdS/CFT correspondence,
which was first proposed by Maldacena in~\cite{Maldacena:1997re,Maldacena:1998re} and states that the strong coupled
field living on the AdS boundary can be described with a weakly gravity theory
in one higher-dimensional AdS spacetime. By applying the AdS/CFT correspondence,
one has first achieved success in the study of holographic QCD and heavy
ions collisions. In recent times, The AdS/CFT correspondence also has provided insights into
condensed matter theory. In \cite{Gubser:2008px,Hartnoll:2008vx,Hartnoll:2008kx}, the authors investigated the action of a
complex scalar field coupled to a $U(1)$ gauge field in a (\(3+1\))-dimensional Schwarzschild-AdS
black hole and found that below some critical value of the temperature due to the
$U(1)$ symmetry breaking, the scalar field  which condenses near the horizon could be
interpreted as a Cooper pair-like superconductor condensation. Moreover, analyzing the
optical conductivity of the superconducting state, the rate of the width of the gap to the
critical temperature is close to the value of a high temperature superconductor. Thus,
it is hoped that the holographic model can match the properties of the high temperature
superconducting behavior. Soon, The p-wave and d-wave holographic superconductors
proposed in \cite{Gubser:2008wv,Chen:2010mk,Benini:2010pr}, respectively. For reviews of holographic superconductors,
see ~\cite{Hartnoll:2009sz,Herzog:2009xv,Horowitz:2010gk,Cai:2015cya}.

It is well known that the Josephson junction is a device made up of two superconductor
materials coupled by weak link barrier in \cite{Josephson:1962zz}. The weak link can be a thin normal
conductor; then it is named a superconductor-normal-superconductor junction
(SNS), or a thin insulating barrier; then we have a superconductor-insulator-superconductor junction (SIS). Recently, by studying the  space-dependent solution of  the action of  a Maxwell field coupled with a complex scalar field in  a (\(3+1\))-dimensional Schwarzschild-AdS black hole background, Horowitz et al.~\cite{Horowitz:2011dz} had  constructed a (\(2+1\))-dimensional holographic model of a Josephson junction and  found a sine relation between the tuning current and the phase difference of the condensation across the junction. The extension to a 4-dimensional Josephson junction has been discussed in \cite{Wang:2011rva,Siani:2011uj}. The  holographic mode  of a Josephson junction array  has been constructed based on a designer multigravity in~\cite{Kiritsis:2011zq}. With the $SU(2)$ gauge field coupled with gravity, the holographic p-wave Josephson junction has also been discussed in~\cite{Wang:2011ri}. In~\cite{Wang:2012yj}, a holographic model of (\(1+1\))-dimensional superconductor-insulator-superconductor (S-I-S) Josephson junction has been investigated.  In \cite{Cai:2013sua,Takeuchi:2013kra}, with the action of the Einstein-Maxwell-complex scalar field, the authors studied a holographic model of superconducting quantum interference device (SQUID). In \cite{Li:2014xia} a holographic model of a Josephson junction in the non-relativistic case with a Lifshitz geometry was constructed.

Recently, the holographic approach has been applied to the coexistence and competition  order phenomena, in which
the phase diagram has a rich structure, such as  the competition of two scalar order parameters in the probe limit~\cite{Basu:2010fa,Musso:2013rnr} and  with the backreaction of the scalars~\cite{Cai:2013wma}, the coexistence of two p-wave order parameters~\cite{Amoretti:2013oia}, and the competition of s-wave and d-wave order parameters~\cite{Li:2014wca}.
Especially, there is another interesting paper  about competition  and coexistence  of s-wave and p-wave order in~\cite{Nie:2013sda,Amado:2013lia}; the authors studied a  charged triplet  scalar coupled with an $SU(2)$ gauge field in the (\(3+1\))-dimensional  spacetime and confirmed
s+p coexisting phases. Furthermore, the phase transitions of the holographic s+p-wave superconductor with backreaction is investigated in Ref.~\cite{Nie:2014qma}.

Since the holographic model of hybrid and s+p coexisting superconductors has been constructed, it is natural to set up a holographic model for hybrid and coexisting s-wave and p-wave Josephson junction. We will study a non-Abelian $SU(2)$ Yang$-$Mills field and a scalar triplet charged under an $SU(2)$ gauge field in (\(3+1\))-dimensional AdS spacetime and construct the holographic model of a hybrid and coexisting s-wave and p-wave Josephson junction, which will be related to the s-N-s junction in~\cite{Horowitz:2011dz} and p-N-p junction in~\cite{Wang:2011ri}. To construct  the holographic model of a junction, ones need to tune the value of the chemical potential.

The paper is organized as follows: in Sect.\ref{sec2}, we consider a non-Abelian $SU(2)$ gauge field coupled with a scalar triplet charged field in (\(3+1\))-dimensional AdS spacetime and construct a gravity dual model for a (\(2+1\))-dimensional s+p coexisting Josephson junction. We show the numerical results of the equations of motion and study the characteristics of the (\(2+1\))-dimensional s+p coexisting Josephson junction in Sect.\ref{sec3}. The conclusion and discussion are in the last section.

\section{Holographic model of hybrid and coexisting s-wave and p-wave Josephson junction}\label{sec2}
\subsection{The model setup}

In~\cite{Nie:2013sda}, the author considered the action of a charged scalar field  coupled to an $SU(2)$ gauge field, in which the charged scalar field
transform as a triplet under the gauge group $SU(2)$. This model can  realize the competition and coexistence  of s-wave and p-wave order parameters. We will also adopt the  same form of the action in the (\(3+1\))-dimensional AdS spacetime:
\be
S=\frac{1}{2 \kappa_g ^2}\int d^{4}x \sqrt{-g} (R-2\Lambda)+\frac{1}{g_c^2}\int d^{4}x \sqrt{-g}(-|D_\mu \Psi^{a}|^{2}-\frac{1}{4}F^a_{\mu\nu}F^{a\mu\nu}-m^2 |\Psi^a|^{2}), \label{maction}
\ee
where $\Lambda=-3/L^{2}$ is the negative cosmology constant, and $L$ is the radius of asymptotic AdS spacetime. $\Psi^{a}$ is a complex scalar triplet charged under the $SU(2)$ Yang$-$Mills field and
\be
 D_{\mu}\Psi^{a}=\partial_{\mu}\Psi^{a}+\varepsilon^{abc}A_{\mu}^{b}\Psi^{c}.
\ee
The field strength $F_{\mu\nu}^{a}$  of the $SU(2)$ gauge theory  is given by
\be
  F_{\mu\nu}^{a} =\partial_{\mu}A_{\nu}^{a}-\partial_{\nu}A_{\mu}^{a}+\varepsilon^{abc}A_{\mu}^{b}A_{\nu}^{c},
\ee
where the one form $A=A_{\mu}dx^{\mu}=A_{\mu}^{a}\tau^{a}dx^{\mu}$  and the superscript a is the index of generator $\tau^{a}$ of $SU(2)$ gauge field with $a=1, 2, 3$. $g_{c}$ is the coupling constant of non-Abelian $SU(2)$ Yang$-$Mills field. In this paper we will consider the probe limit by taking $g_{c}\rightarrow\infty$ to ignore the backreaction of the matter.

In the probe limit, we still choose a (\(3+1\))-dimensional planar Schwarzschild$-$AdS  black hole solution as the  background geometry  with metric
\be \label{metric}
  ds^{2} = -f(r)dt^{2}+\frac{dr^{2}}{f(r)}+r^{2}(dx^{2}+dy^{2}),
\ee
where $x$ and $y$ are the coordinates of a 2-dimensional Euclidean space. The function $f(r)$ is
\be
f(r)=\frac{r^{2}}{L^{2}}(1-\frac{r_{h}^{3}}{r^{3}}),
\ee
where $r_{h}$ is the radius of the black hole's event horizon. The Hawking temperature of the black hole is given by
\be\label{T}
T=\frac{1}{4\pi}\frac{df}{dr}\bigg|_{r=r_{h}}=\frac{3r_{h}}{4\pi L^{2}}.
\ee
The temperature $T$ relates to the radius of the black hole's event horizon $r_{h}$ and the AdS radius $L$. In the rest of paper, we will work in unit in which $L=1$. $T$ corresponds to the temperature of the dual field theory on the AdS boundary.

The variation of the action (\ref{maction}) with respect to the scalar field $\Psi^{a}$ and $A_{\mu}^{a}$ lead to the equations of motion, respectively. We have

\begin{align}
\nabla_{\mu}(D^{\mu}\Psi^{a})+\varepsilon^{abc}A^{b}_{\mu}D^{\mu}\Psi^{c}-m^{2}\Psi^{a}&=0\;,\label{psia}\\
\nabla^{\nu}F^{a}_{\nu\mu}+\varepsilon^{abc}A^{b\nu}F^{c}_{\nu\mu}-2\varepsilon^{abc}\Psi^{b}(\partial_{\mu}\Psi^{c}+\varepsilon^{cde}A^{d}_{\mu}\Psi^{e})&=0\;.\label{amu}
\end{align}

For the hybrid and coexisting s-wave and p-wave Josephson junction, there are several different kinds of
matter fields ansatzes. Let us consider one of them
\be\label{ansz}
\widetilde{\Psi}^{3}=\widetilde{\Psi}_{3}(r,y),\quad A_{t}^{1}=\phi(r,y),\quad A_{x}^{3}=\Psi_{x}(r,y),\quad A_{r}^{1}=A_{r}(r,y),\quad A_{y}^{1}=A_{y}(r,y),
\ee
where the field functions $\widetilde{\Psi}^{3}$, $\theta^{a}$, $\phi$, $\Psi_{x}$, $A_{r}$, and $A_{y}$ are dependent of the coordinates  $r$ and $y$. Thus the holographic model of the Josephson junction would be along the $y$ direction. Without loss of generality, we will consider the $SU(2)$ gauge-invariant vector field $A^{a}_{\mu}$ and the scalar field $\widetilde{\Psi}^{a}$:\\
\be
M^{a}_{\mu}=A_{\mu}^{a}-D_{\mu}\theta^{a},\quad \Psi^{a}=\widetilde{\Psi}^{a}+\varepsilon^{abc}\theta^{b}\widetilde{\Psi}^{c}.
\ee
With the black hole background (\ref{metric}) and the above ansatz (\ref{ansz}), the equations of the matter fields (\ref{psia}) and (\ref{amu}) can be  written as
\begin{subequations}
\begin{align}
\partial_{r}^{2}\Psi_{3}+\frac{1}{r^{2}f}\partial^{2}_{y}\Psi_{3}+(\frac{f'}{f}+\frac{2}{r})\partial_{r}\Psi_{3}
+(-\frac{m^{2}}{f}+\frac{\phi^{2}}{f^{2}}-M_{r}^{2}-\frac{M_{y}^{2}}{r^{2}f})\Psi_{3}&=0\;,\label{psi3}\\
\partial_{r}^{2}\phi+\frac{1}{r^{2}f}\partial_{y}^{2}\phi+\frac{2}{r}\partial_{r}\phi
-\frac{\phi\Psi_{x}^{2}}{r^{2}f}-\frac{2\phi\Psi_{3}^{2}}{f}&=0\;,\label{phi}\\
\partial_{r}^{2}\Psi_{x}+\frac{1}{r^{2}f}\partial_{y}^{2}\Psi_{x}+\frac{f'}{f}\partial_{r}\Psi_{x}
+(\frac{\phi^{2}}{f^{2}}-M_{r}^{2}-\frac{M_{y}^{2}}{r^{2}f})\Psi_{x}&=0\;,\label{psix}\\
\partial_{y}^{2}M_{r}-\partial_{r}\partial_{y}M_{y}-M_{r}\Psi_{x}^{2}-2r^{2}M_{r}\Psi_{3}^{2}&=0\;,\label{Ar}\\
\partial_{r}^{2}M_{y}-\partial_{r}\partial_{y}M_{r}+\frac{f'}{f}(\partial_{r}M_{y}-\partial_{y}M_{r})
-\frac{M_{y}\Psi_{x}^{2}}{r^{2}f}-\frac{2\Psi^{2}_{3}M_{y}}{f}&=0\;,\label{Ay}\\
\partial_{r}M_{r}+\frac{1}{r^{2}f}\partial_{y}M_{y}+\frac{2}{\Psi_{3}}(M_{r}\partial_{r}\Psi_{3}+\frac{M_{y}}{r^{2}f}\partial_{y}\Psi_{3})
+(\frac{f'}{f}+\frac{2}{r})M_{r}&=0\;,\label{Ac1}\\
\partial_{r}M_{r}+\frac{1}{r^{2}f}\partial_{y}M_{y}+\frac{2}{\Psi_{x}}(M_{r}\partial_{r}\Psi_{x}
+\frac{M_{y}}{r^{2}f}\partial_{y}\Psi_{x})+\frac{f'}{f}M_{r}&=0\;,\label{Ac}
\end{align}
\end{subequations}
where a prime denotes the derivative with respect to $r$.  In our paper, we will work with the case  $m^{2}\geq-9/4$ in order to  satisfy the BF bound~\cite{Breitenlohner:1982bm}. Let us inspect Eqs. (\ref{psi3})$-$(\ref{Ac}). If $\Psi_{x}$, $A_{r}$, and $A_{y}$ are turned off, $\Psi_{3}$ and $\phi$ are only dependent on $r$, and the remaining two equations will be the same as the equations of the s-wave holographic superconductivity. Similarly, if we turn off $\Psi_{3}$, $A_{r}$ and $A_{y}$, $\Psi_{x}$, and $\phi$ are only dependent on $r$, and the two  remaining equations  will be the equations which describe the p-wave condensation. Next, we will consider the Josephson junction. If $\Psi_{x}$ is only turned off, the remaining equations are the same equations as of the pure s-wave Josephson junction. In a similar way, if we only turn off $\Psi_{3}$, we will get the pure p-wave Josephson junction. So we can get the so-called s+p coexisting phase Josephson junction under this ansatz (\ref{ansz}). It is obvious that  Eqs. (\ref{psi3})$-$(\ref{Ac}) are coupled and nonlinear, so we need to solve them numerically instead of solving them analytically.

Near the AdS boundary ($r\rightarrow\infty$), the matter fields take the asymptotic forms
\begin{align}
\Psi_{3}&=\frac{\Psi_{3}^{(-)}(y)}{r^{\Delta_{-}}}+\frac{\Psi_{3}^{(+)}(y)}{r^{\Delta_{+}}}+...\;,\\
\phi&=\mu(y)-\frac{\rho(y)}{r}+\mathcal{O}\left(\frac{1}{r^2}\right)\;,\\
\Psi_{x}&=\Psi_{x}^{(-)}(y)+\frac{\Psi_{x}^{(+)}(y)}{r}
+\mathcal{O}(\frac{1}{r^{3}})\;,\\
M_r&=\mathcal{O}\left(\frac{1}{r^3}\right)\;,\\
M_y&=\nu(y)+\frac{J}{r}+\mathcal{O}\left(\frac{1}{r^2}\right)\;. \label{defnu}
\end{align}
Here, the dimensions of the operations $\Psi_{3}^{(\pm)}$ are
\be
\Delta_{\pm}=\frac{3\pm\sqrt{9+4m^{2}}}{2}.
\ee
According to the AdS/CFT dictionary,
$\Psi_{3}^{(\pm)}$ is considered as the source of the scalar operation of s-wave condensation, and $\Psi_{3}^{(\mp)}$ is the corresponding expectation value of the operator. Meanwhile,
$\Psi_{x}^{(\pm)}$ is the source of the vector operation of p-wave condensation, and $\Psi_{x}^{(\mp)}$ is the corresponding expectation value of the operator. $\mu(y)$, $\rho(y)$, $\nu(y)$, and $J$  are  the chemical potential, charge density, velocity of the superfluid, and the constant current in the dual field, respectively~\cite{Basu:2008st,Herzog:2008he,Arean:2010xd,Sonner:2010yx,Horowitz:2008bn,Arean:2010zw,Zeng:2010fs,Arean:2010wu}.

In order to solve Eqs. (\ref{psi3})$-$(\ref{Ac}) numerically, we need to impose the boundary conditions on them. First, we impose the Dirichlet boundary condition on $\Psi_{3}$ and $\Psi_{x}$ on the AdS boundary. In this paper, we set
\be
\Psi_{3}^{(-)}=0.
\ee
So $\Psi_{3}^{(+)}$ is the expectation value of s-wave scalar operator, $\Psi_{3}^{(+)}=\langle\mathcal{O}_{s}\rangle$. Similarly, we impose the Dirichlet boundary condition on $\Psi_{x}$ on the AdS boundary:
\be
\Psi_{x}^{(-)}=0.
\ee
$\Psi_{x}^{(+)}$ is the expectation value of the p-wave vector operator, $\Psi_{x}^{(+)}=\langle\mathcal{O}_{p}\rangle$. Second, we impose the Dirichlet boundary condition on $\phi$ at the horizon:
\be
\phi(r_{h})=0,
\ee
$\phi$ is the $t$ component of $A_{\mu}^{1}$, and $\phi(r_{h})=0$ is to avoid the divergence of $g^{\mu\nu}A_{\mu}^{1}A_{\nu}^{1}$. In addition, the matter field functions are independent of $y$ at the spatial coordinate $y\rightarrow\pm\infty$. So, the boundary conditions of the Eqs. (\ref{psi3})$-$(\ref{Ac}) are determined by the value of $\mu$ and $J$. The phase difference $\gamma$ across the junction is $\gamma=\Delta\theta^{1}-\int A_{y}dy$. The phase difference $\Delta\theta^{1}$ can be eliminated under the $SU(2)$ gauge-invariant, thus it is convenient to set
\be
\gamma=-\int_{-\infty}^{+\infty} A_{y}dy.
\label{gamma}
\ee

Now, we would like to introduce the critical temperature $T_{c}$ of the Josephson junction, which is proportional to the chemical potential
$\mu(+\infty)$ or $\mu(-\infty)$, and  we set
\be\label{T_{c}}
T_{c}=\frac{3}{4\pi}\frac{\mu(-\infty)}{\mu_{c}}\;
\ee
where $\mu_{c}\approx 3.65$.
In order to describe a hybrid and coexisting s-wave and p-wave Josephson junction, we should construct a chemical potential which can make a phase transition  occuring along the direction $y$ of  the Josephson junction; thus the chemical potential $\mu(y)$ is dependent on the spatial coordinate $y$,
\be\label{profile}
\mu(y)=\mu\left\{1-\frac{1-\epsilon}{2\tanh(\frac{\ell}{2\sigma})}\left[\tanh\left(\frac{y+\tfrac{\ell}{2}}{\sigma}\right)-\beta\tanh\left(\frac{y-\tfrac{\ell}{2}}{\sigma}\right)\right]\right\}\;,
\ee
where the chemical potential $\mu(y)$ is proportional to $\mu$ and $\ell$ is the width of the Josephson junction. We use the parameters $\epsilon$, and $\sigma$ and $\beta$ control the steepness and the depth of the junction, respectively.

\subsection{The scaling symmetry}
Analyzing the EoM, we found that there is a scaling symmetry in Eqs. (\ref{psi3})$-$(\ref{Ac}). These equations are invariant under the following scale transformation:
\begin{displaymath}\label{Bscaling}
\left\{ \begin{array}{ll}
r\rightarrow br,\quad r_{h}\rightarrow br_{h},\quad y\rightarrow y/b,\quad \phi\rightarrow b\phi,\qquad\qquad\qquad\qquad
\nonumber\\
\Psi_{x}\rightarrow b\Psi_{x},\quad \Psi_{3}\rightarrow\Psi_{3},\quad f\rightarrow b^{2}f,\quad f'\rightarrow bf',\quad A_{y}\rightarrow bA_{y}.
\end{array} \right.
\end{displaymath}
Under this scaling symmetry, we can get the behavior of the following physical quantities:
\begin{displaymath}
\left\{ \begin{array}{ll}
\mu\rightarrow b\mu,\quad\Psi_{3}^{(-)}\rightarrow b^{\Delta_{-}}\Psi_{3}^{(-)},\quad \Psi_{3}^{(+)}\rightarrow b^{\Delta_{+}}\Psi_{3}^{(+)},\qquad\quad
\nonumber\\
\rho\rightarrow b^{2}\rho,\quad\Psi_{x}^{(-)}\rightarrow
b\Psi_{x}^{(-)},\quad
\Psi_{x}^{(+)}\rightarrow
b^{2}\Psi_{x}^{(+)},\quad J\rightarrow b^{2}J.
\end{array} \right.
\end{displaymath}
Because of this scaling symmetry, we can set the radius of the black hole's event horizon $r_{h}=1$. In addition, we have set $L=1$, so the temperature $T$ and the background geometry is fixed. From Eq. (\ref{T}), we can see that the temperature $T$ changes to $bT$ under the scaling transformation. So the following quantities are invariant:
\be
\Psi_{3}^{(+)}/T^{\Delta_{+}},\quad\Psi_{x}^{(+)}/T^{2},\quad J/T^{2}.\nonumber
\ee
These invariant quantities can change with $T/T_{c}$.

\section{Numerical results}\label{sec3}
In this section we will solve the coupled and nonlinear equations (\ref{psi3})$-$(\ref{Ac}) numerically. Before we solve these equations, we will have coordinate transformations, $z=1/r$, $\widetilde{y}=\tanh(\frac{y}{4\sigma})$. It is more convenient to impose boundary conditions at $z=1$, $z=0$, and $\widetilde{y}=\pm1$, rather than at $r=1$, $r=+\infty$, and $y=\pm\infty$. There are four kinds of Josephson junctions.\\
(i)The s+p-N-s+p Josephson junction consists of the s+p coexisting phase in the two leads, and the normal phase between them.\\
(ii)The s+p-N-s Josephson junction consists of the s+p coexisting phase in the left lead, the conventional s-wave phase in the right lead, and the normal phase between them.\\
(iii)The s+p-N-s Josephson junction consists of the s+p coexisting phase in the left lead, the conventional p-wave phase in the right lead, and the normal phase between them.\\
(iv)The s-N-p Josephson junction consists of the conventional s-wave phase in the left lead, the conventional s-wave phase in the right lead, and the normal phase between them.

We can tune the chemical potential $\mu(y)$ to realize the four cases. In~\cite{Nie:2013sda}, when the value of $\Delta_+$ or $m^{2}$ is in a special region, as the temperature decreases, the p-wave condensation will appear first and increase; the s-wave condensation will not appear. When the temperature continues to decrease and reaches the critical temperature $T_{c1}^{sp}$, the s-wave condensation will appear and increase, and the p-wave condensation will decrease. When the temperature reaches another critical temperature $T_{c2}^{sp}$, the p-wave condensation will disappear. So when the temperature is in the region $T_{c2}^{sp}\sim T_{c1}^{sp}$, the s-wave phase and p-wave will coexist. When the temperature is higher than $T_{c1}^{sp}$, there is only a p-wave phase. When the temperature is blew $T_{c2}^{sp}$, there is only an s-wave phase. In order to construct the junction of the s+p coexisting phase, we can write the region of the chemical potential as  $\mu_{c1}\sim\mu_{c2}$ in \cite{Nie:2013sda}.
\subsection{s+p-N-s+p Josephson junction}
\begin{figure}
  \begin{center}
  \subfigure{
  \includegraphics[width=0.43\textwidth,height=0.28\textheight]{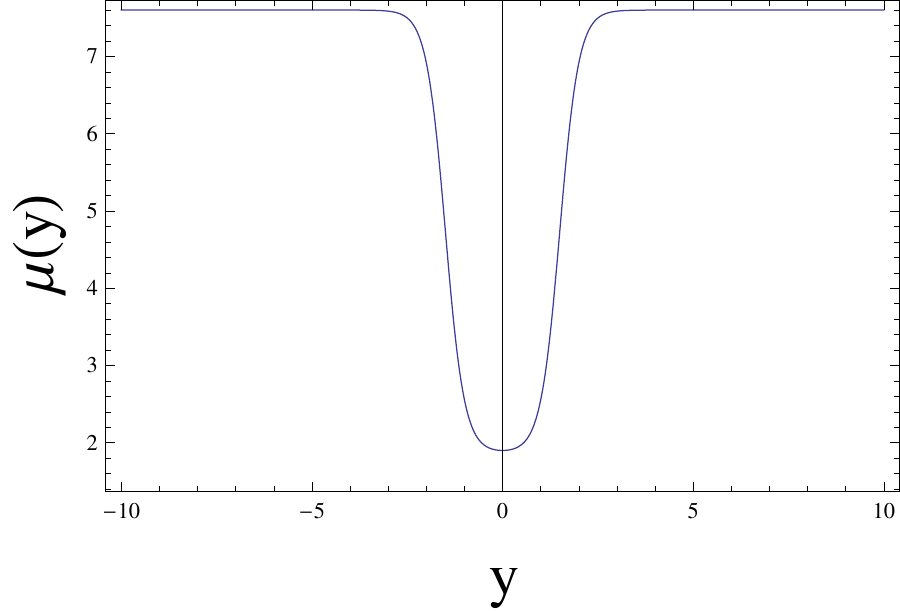}}
\hspace{0.20cm}
 \subfigure{
  \includegraphics[width=0.51\textwidth,height=0.3\textheight]{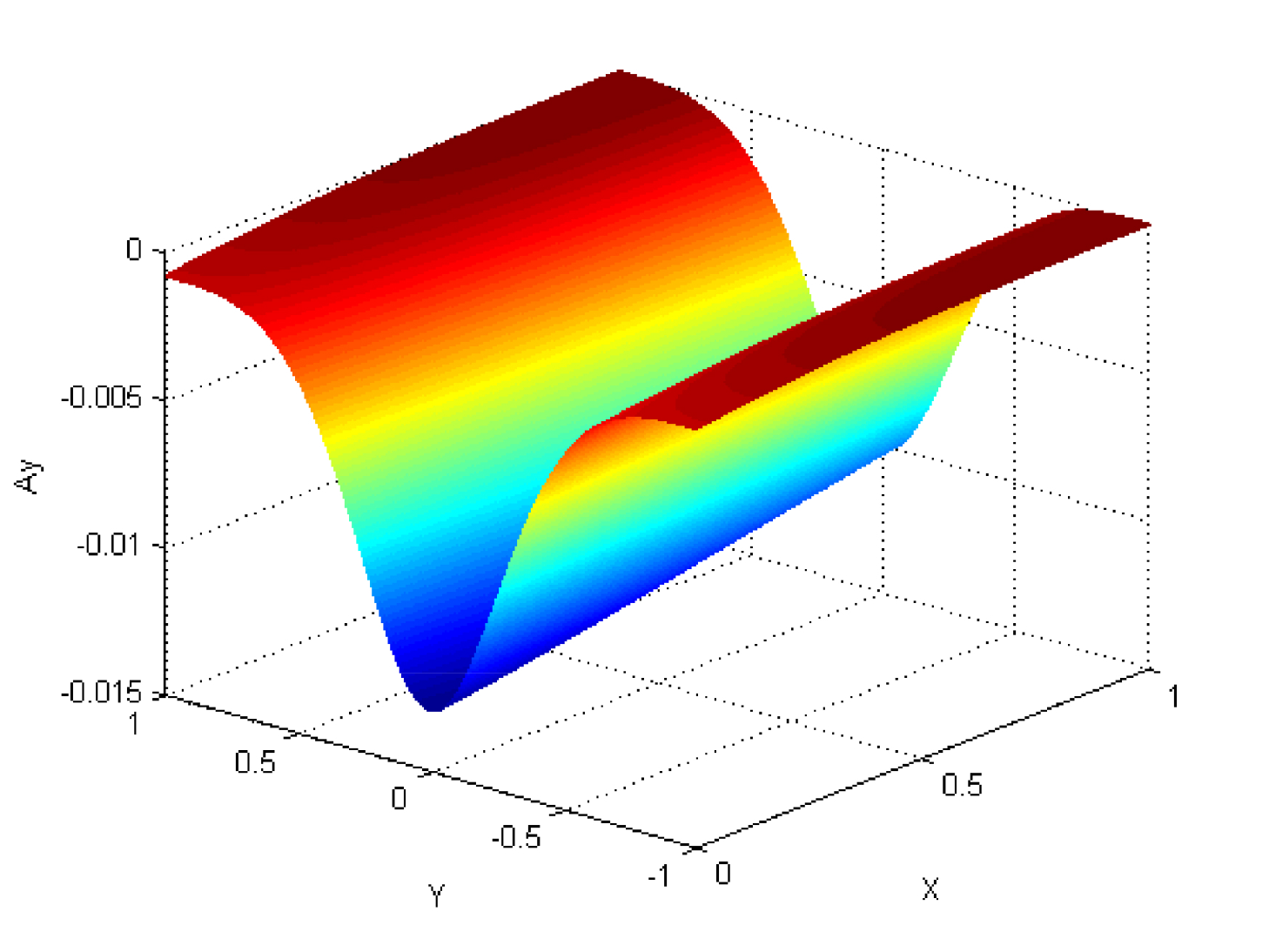}}
\end{center}  \vskip -5mm
  \caption{We represent the components $\mu(y)$ and $A_{y}$ of the Yang$-$Mills field. \emph{Left} The profile of $\mu(y)$. $\emph{Right}$ The figure of $A_{y}$. The parameters are $m^{2}=-33/16$, $\mu=7.6$, $\epsilon=0.25$, $\sigma=0.5$, $\ell=3$, and $\beta=1$. We can see that these two figures have symmetry.}
\label{fig_scalar_warpfactor1}
\end{figure}

In this subsection, in order to  obtain the model of a s+p-N-s+p Josephson junction, we need  to tune the value of chemical potential $\mu(y)$ at $y=\pm\infty$ such that it is in the s+p coexisting region $\mu_{c1}\sim\mu_{c2}$. Because  the superconductor phase in the two leads are symmetrical, the phase difference $\gamma$ can be obtained by (\ref{gamma}); we have
\be
\gamma=-\int^{+\infty}_{-\infty}dy[\nu(y)-\nu(\pm\infty)].
\ee
The profiles of $A^{1}_{t}$ and $A_{y}^{1}$ are shown in Fig. \ref{fig_scalar_warpfactor1}.

\begin{figure}
  \begin{center}
  \subfigure{
  \includegraphics[width=0.47\textwidth,height=0.28\textheight]{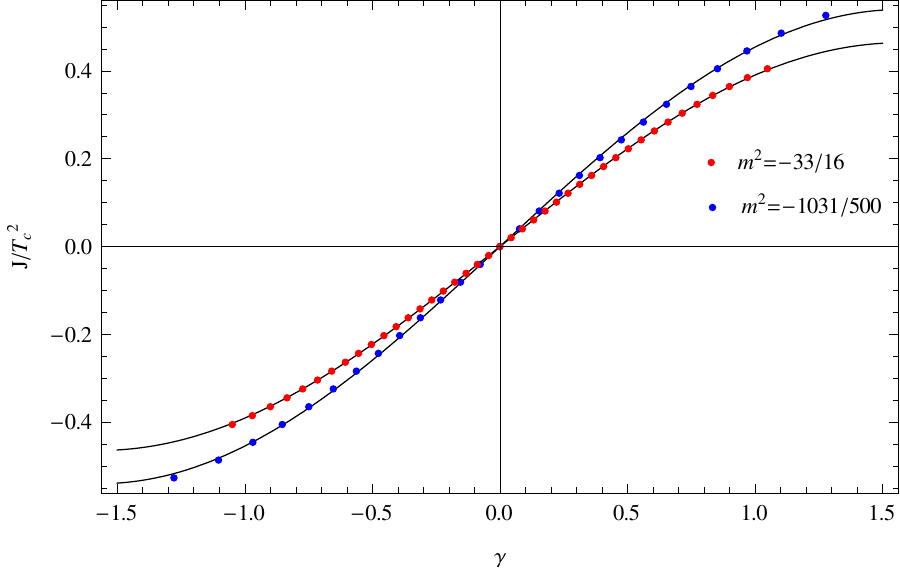}}
\hspace{0.20cm}
 \subfigure{
  \includegraphics[width=0.47\textwidth,height=0.28\textheight]{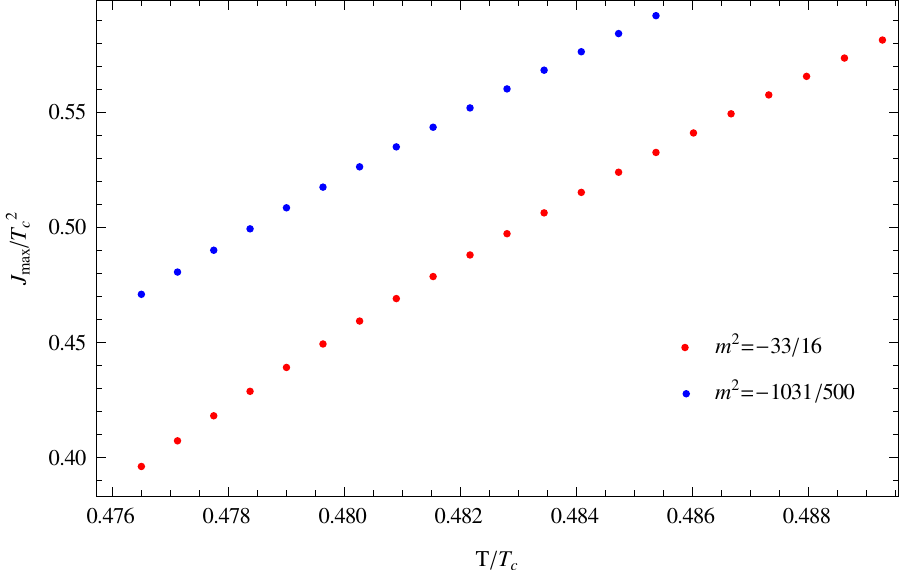}}
\end{center}  \vskip -5mm
  \caption{\emph{Left} The current $J/T_{c}^{2}$ as a sine function of the difference phase $\gamma$. \emph{Right} The $J_{max}/T_{c}^{2}$ increases with growing $T/T_{c}$. The black lines are sine curves and the points with $m^{2}=-33/16$(\emph{red}), $-1031/500$(\emph{blue}) are numerical results. The parameters are $\mu=7.6$, $\epsilon=0.25$, $\sigma=0.5$, $\ell=3$, and $\beta=1$.}
\label{fig_jgammaspnsp jmaxtspnsp}
\end{figure}

We show the relationship between the DC current $J$ and the phase difference $\gamma$ across the junction on the left panel of Fig. \ref{fig_jgammaspnsp jmaxtspnsp}. From the figure, we can see that $J$ is proportional to the sine of $\gamma$. We have
\begin{displaymath}
J/T_{c}^{2}\approx\left\{ \begin{array}{ll}
 0.464sin\gamma & \qquad\textrm{for \quad$m^{2}=-33/16$},\\
 0.539sin\gamma & \qquad\textrm{for \quad$m^{2}=-1031/500$}.
 \end{array} \right.
\end{displaymath}
Note that we only obtain the phase difference in the interval ($-\pi/2,\pi/2$), and we can see that the points which represent our numerical data  fit the sine line very well. We can see that when the value of $m^{2}$ increases, the maximum current $J_{max}$ will grow. The dependence of $J_{max}$ on the temperature $T$ is shown on the right panel of Fig. \ref{fig_jgammaspnsp jmaxtspnsp}. The graph shows that $J_{max}/T_{c}^{2}$ increases with growing $T/T_{c}$, however, in the s-N-s or p-N-p Josephson junction \cite{Horowitz:2011dz,Wang:2011ri}, $J_{max}/T_{c}^{2}$ decays with increasing $T/T_{c}$. The reason is that in the s-wave and p-wave coexisting region \cite{Nie:2013sda}, the condensation of the s-wave decreases and the condensation of the p-wave increases with growing temperature, respectively. But the total condensation would  decrease when the temperature drops. So, the $J_{max}$ decreases with rising $T$.

\begin{figure}
  \begin{center}
  \subfigure{
  \includegraphics[width=0.47\textwidth,height=0.28\textheight]{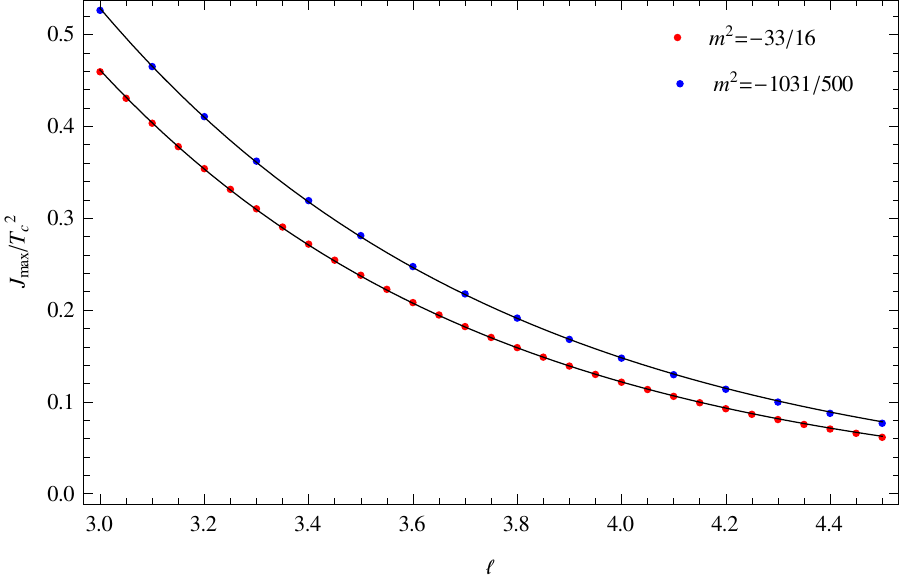}}
\hspace{0.20cm}
 \subfigure{
  \includegraphics[width=0.47\textwidth,height=0.28\textheight]{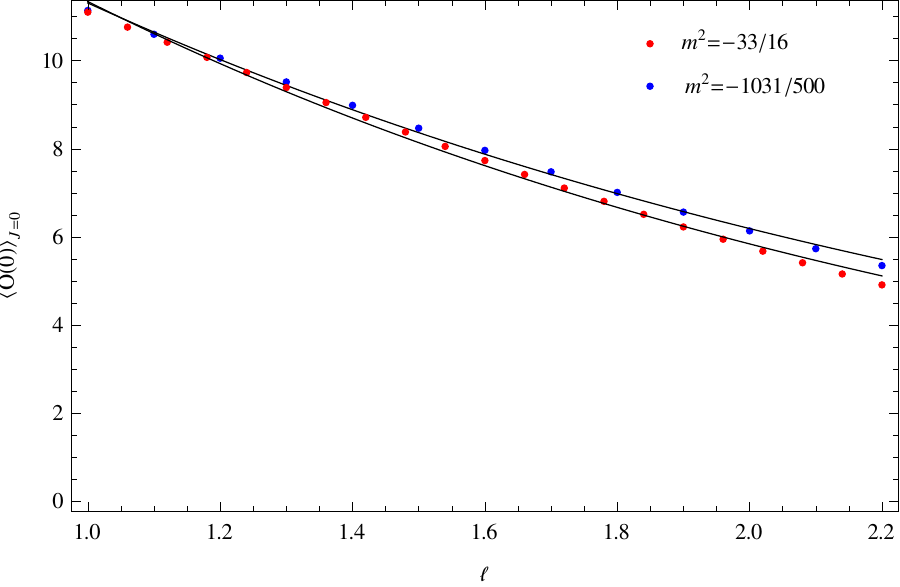}}
\end{center}  \vskip -5mm
  \caption{\emph{Left} The maximum current $J_{max}$ as exponential function of $\ell$. \emph{Right} The total condensate $\langle\mathcal{O}(0)\rangle_{J=0}$ as exponential function of $\ell$. Our numerical results are the points with $m^{2}=-33/16$(\emph{red}), $-1031/500$(\emph{blue}). In all the plots, we use $\mu=7.6$, $\epsilon=0.25$, $\sigma=0.5$, and $\beta=1$. The numerical data fit the exponential curves well.}
\label{fig_jamxlspnsp olspnsp}
\end{figure}

The relationship of between $J_{max}/T_{c}^{2}$ and $\ell$ is shown on the left panel of Fig. \ref{fig_jamxlspnsp olspnsp}. The $J_{max}/T_{c}^{2}$ will decay with $\ell$  exponentially and the change will be larger when $m^{2}$ becomes larger. The total condensation in $y=0$ when the current $J=0$ is the sum of s-wave and p-wave condensation. Here, we define
\be
\langle\mathcal{O}(0)\rangle_{J=0}=\langle\mathcal{O}_{s}(0)\rangle_{J=0}/T_{c}^{\Delta}+\langle\mathcal{O}_{p}(0)\rangle_{J=0}/T_{c}^{2}.
\ee
We plot the $\langle\mathcal{O}(0)\rangle_{J=0}$ on the right panel in Fig. \ref{fig_jamxlspnsp olspnsp}. $\langle\mathcal{O}(0)\rangle_{J=0}$ also decays with growing $\ell$ exponentially and $\langle\mathcal{O}(0)\rangle_{J=0}$ becomes larger with increasing $m^{2}$. We have
\begin{displaymath}
\left\{ \begin{array}{ll}
 J_{max}/T_{c}^{2}\approx25.12e^{-\ell/0.7505} \\
 \qquad\qquad\qquad\qquad  & \qquad\textrm{for \quad$m^{2}=-33/16$}.\\
 \langle\mathcal{O}(0)\rangle_{J=0}\approx22.03e^{-\ell/(2\times0.7534)}
 \end{array} \right.
\end{displaymath}
We can obtain the coherence length (0.7505, 0.7534) from the above two equations, respectively. The error of the two values is about 0.4 \%. We have
\begin{displaymath}
\left\{ \begin{array}{ll}
 J_{max}/T_{c}^{2}\approx24.08e^{-\ell/0.7856} \\
 \qquad\qquad\qquad\qquad  & \qquad\textrm{for \quad$m^{2}=-1031/500$}.\\
 \langle\mathcal{O}(0)\rangle_{J=0}\approx20.66e^{-\ell/(2\times0.8295)}
 \end{array} \right.
\end{displaymath}
The coherence length (0.7856, 0.8295) is obtained from the above two equations, respectively. The error of the two values is about 5.6 \%.

\subsection{s+p-N-s Josephson junction}
\begin{figure}
  \begin{center}
  \subfigure{
  \includegraphics[width=0.47\textwidth,height=0.28\textheight]{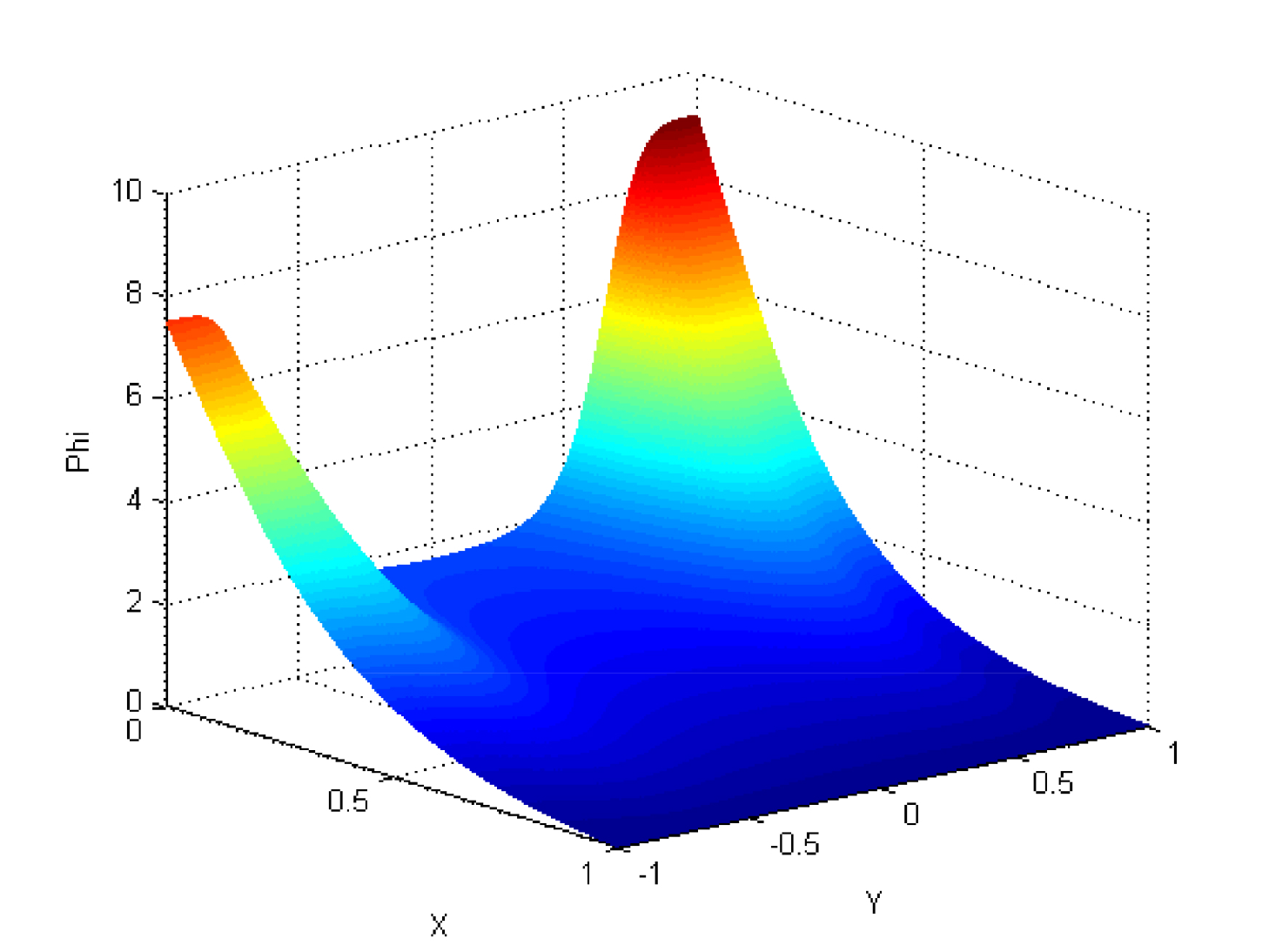}}
\hspace{0.2cm}
 \subfigure{
  \includegraphics[width=0.47\textwidth,height=0.28\textheight]{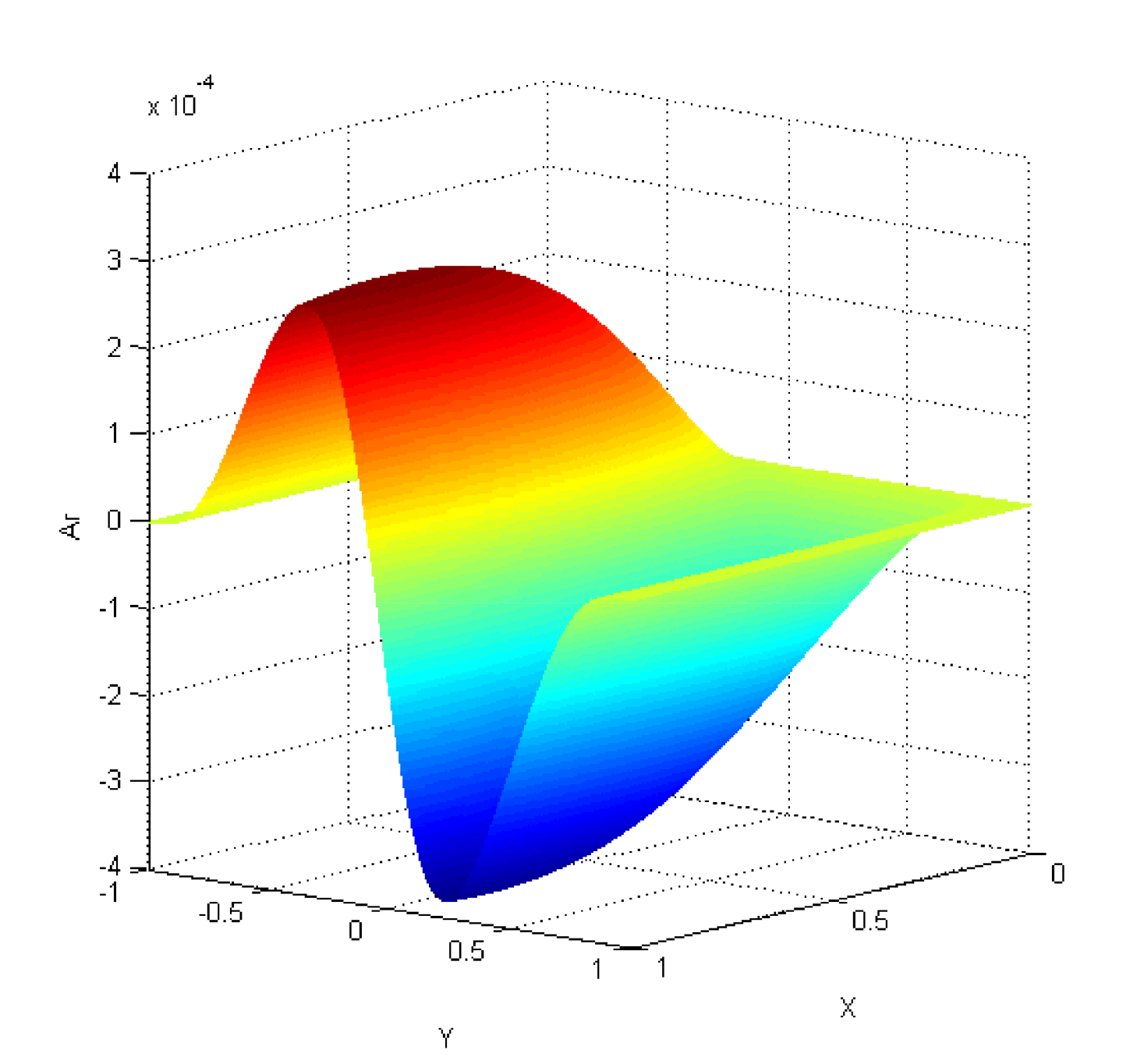}}
\end{center}  \vskip -5mm
  \caption{We represent the components $\phi$ and $A_{r}$ of the Yang-Mills field. \emph{Left} The profile of $\phi$. \emph{Right} The figure of $A_{r}$. The parameters are $m^{2}=-33/16$, $\mu=8.3$, $\epsilon=0.3$, $\sigma=0.5$, $\ell=3$, and $\beta=1.286$. We can see that these two figures are not symmetrical.}
\label{fig_scalar_warpfactor2}
\end{figure}

In this subsection, the model of  the s+p-N-s Josephson junction will be constructed. We tune the chemical potential $\mu(y)$ such that $\mu(-\infty)$ is in the s+p coexisting region $\mu_{c1}\sim\mu_{c2}$ and $\mu(+\infty)>\mu_{c2}$ is in the pure s-wave phase region. Because the superconductor condensations in the two leads are not symmetrical, the phase difference $\gamma$  can be calculated by (\ref{gamma})
\be\label{gamma1}
\gamma=-\int_{-\infty}^{0}dy[\nu(y)-\nu(-\infty)]-\int_{0}^{+\infty}dy[\nu(y)-\nu(+\infty)].
\ee
First, we show the profiles of $A^{1}_{t}$ and $A_{y}^{1}$ in Fig. \ref{fig_scalar_warpfactor2}, with the parameters $m^{2}=-33/16$, $\mu=8.3$, $\epsilon=0.3$, $\sigma=0.5$, $\ell=3$, and $\beta=1.286$.

\begin{figure}
  \begin{center}
  \subfigure{
  \includegraphics[width=0.47\textwidth,height=0.28\textheight]{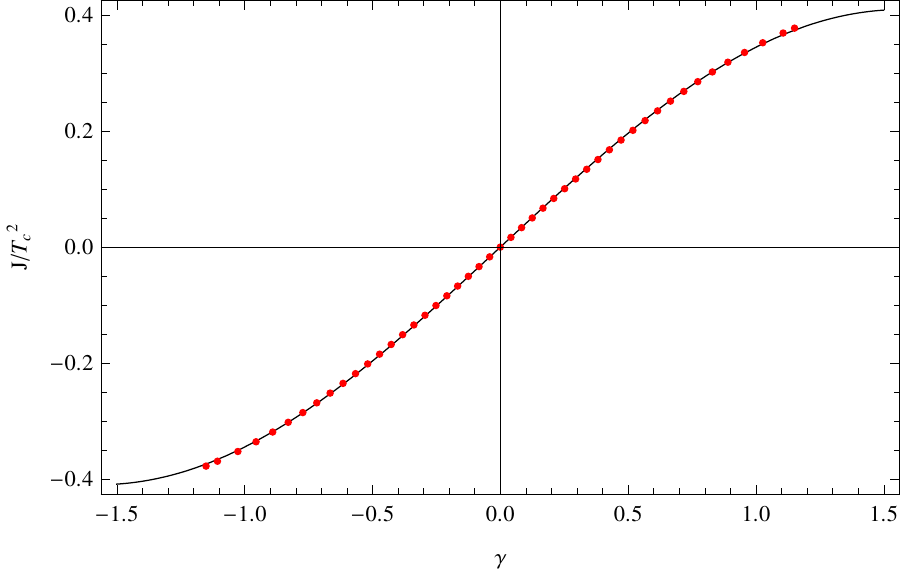}}
\hspace{0.2cm}
 \subfigure{
  \includegraphics[width=0.47\textwidth,height=0.28\textheight]{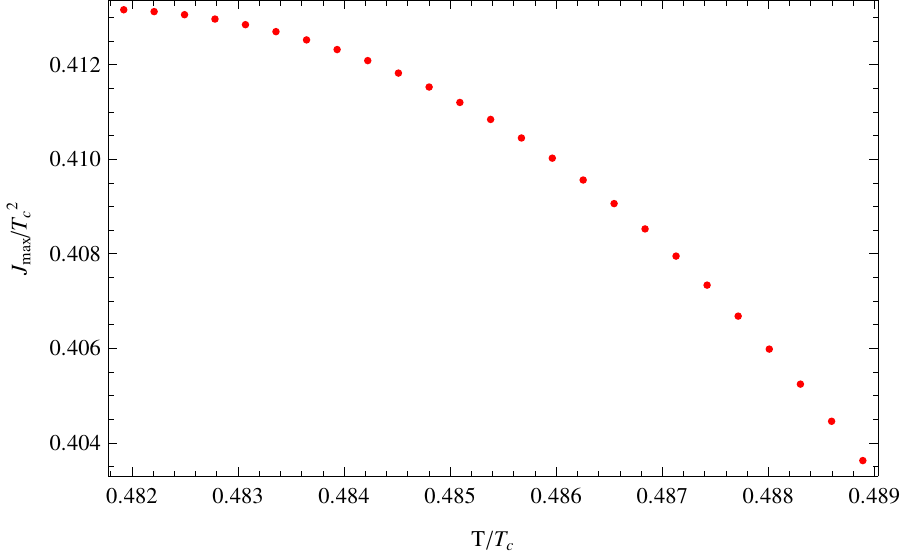}}
\end{center}  \vskip -5mm
  \caption{\emph{Legt} The current $J/T_{c}^{2}$ as sine function of phase difference $\gamma$. \emph{Right} The $J_{max}/T_{c}^{2}$ decays with growing $T/T_{c}$. The parameters are $m^{2}=-33/16$, $\mu=8.3$, $\epsilon=0.3$, $\sigma=0.5$, $\ell=3$ and $\beta=1.286$. The numerical data fit the exponential curves well.}
\label{fig_jgammaspns jmaxtspns}
\end{figure}

The dependence of the current $J$ on the phase difference $\gamma$ across the junction is shown on the left panel of Fig. \ref{fig_jgammaspns jmaxtspns}. The figure shows that $J$ is proportional to the sine of $\gamma$.

\be
J/T_{c}^{2}\approx0.4095sin\gamma,\qquad for\quad m^{2}=-33/16.
\ee
From the above result, we can see our numerical data which is drawn with the red points fits sine line very well. The dependence of $J_{max}$ on the temperature $T$ is shown on the right panel of Fig. \ref{fig_jgammaspns jmaxtspns}. The graph shows that $J_{max}/T_{c}^{2}$ decreases with growing $T/T_{c}$.

\begin{figure}
  \begin{center}
  \subfigure{
  \includegraphics[width=0.47\textwidth,height=0.28\textheight]{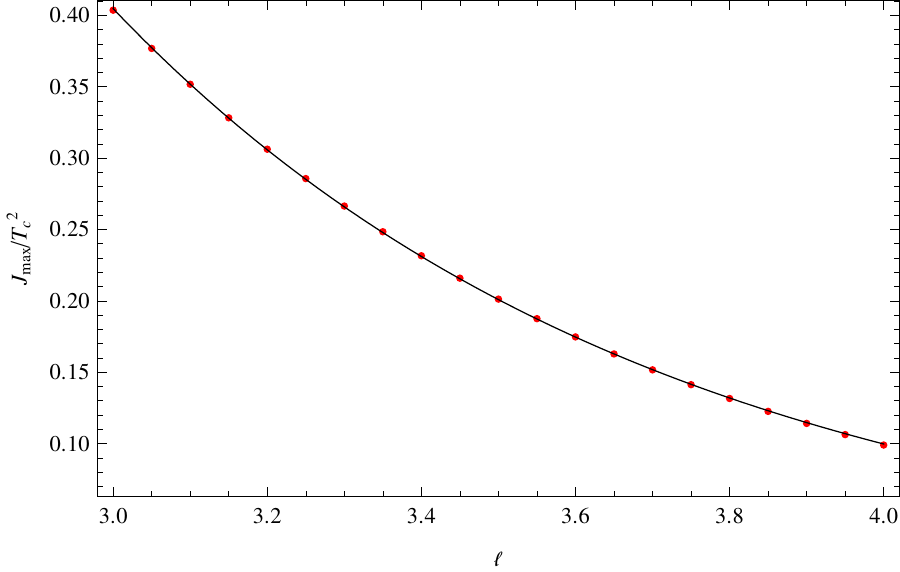}}
\hspace{0.2cm}
 \subfigure{
  \includegraphics[width=0.47\textwidth,height=0.28\textheight]{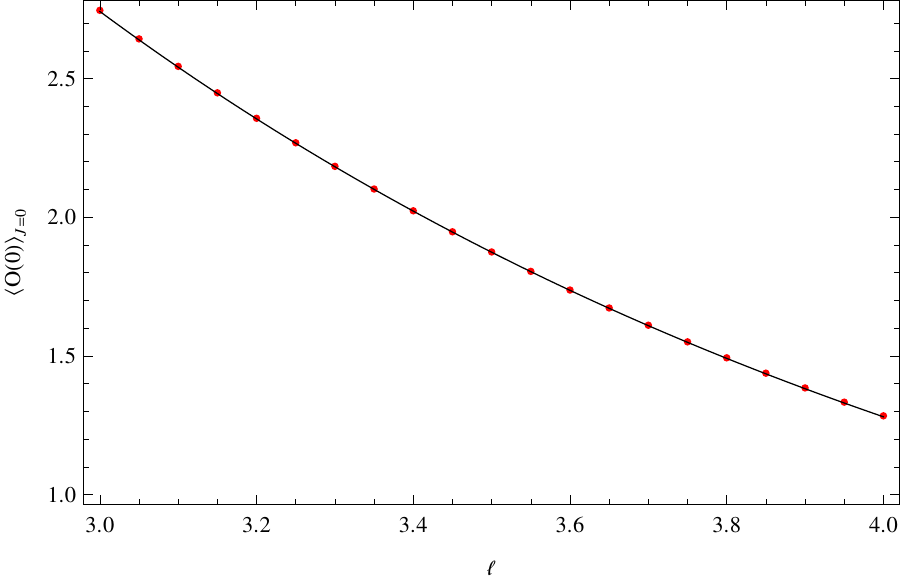}}
\end{center}  \vskip -5mm
  \caption{\emph{Left} The maximum current $J_{max}/T_{c}^{2}$ as an exponential function of the width $\ell$. \emph{Right} The total condensate $\langle\mathcal{O}(0)\rangle_{J=0}$ as an exponential function of the width $\ell$. In all the plots, we use $m^{2}=-33/16$, $\mu=8.3$, $\epsilon=0.3$, $\sigma=0.5$ and $\beta=1.286$. The numerical data fit exponential curves well.}
\label{fig_jamxlspns olspns}
\end{figure}

The dependence of $J_{max}$  on the width $\ell$ of the gap is shown on the left panel of Fig. \ref{fig_jamxlspns olspns}. The figure shows that $J_{max}/T_{c}^{2}$ decays with growing $\ell$ exponentially. The dependence of $\langle\mathcal{O}(0)\rangle_{J=0}$ on the width $\ell$ of the gap is shown on the right panel of Fig. \ref{fig_jamxlspns olspns}. The figure predicts that $\langle\mathcal{O}(0)\rangle_{J=0}$ decays with growing $\ell$ exponentially. We have
\begin{displaymath}
\left\{ \begin{array}{ll}
 J_{max}/T_{c}^{2}\approx27.07e^{-\ell/0.7138} \\
 \qquad\qquad\qquad\qquad  & \qquad\textrm{for \quad$m^{2}=-33/16$}.\\
 \langle\mathcal{O}(0)\rangle_{J=0}\approx26.91e^{-\ell/(2\times0.6567)}
 \end{array} \right.
\end{displaymath}
The coherence length (0.7138, 0.6567) can be obtained from these equations. The error of these two values is about 8 \%.

\subsection{s+p-N-p Josephson junction}

\begin{figure}
  \begin{center}
  \subfigure{
  \includegraphics[width=0.47\textwidth,height=0.28\textheight]{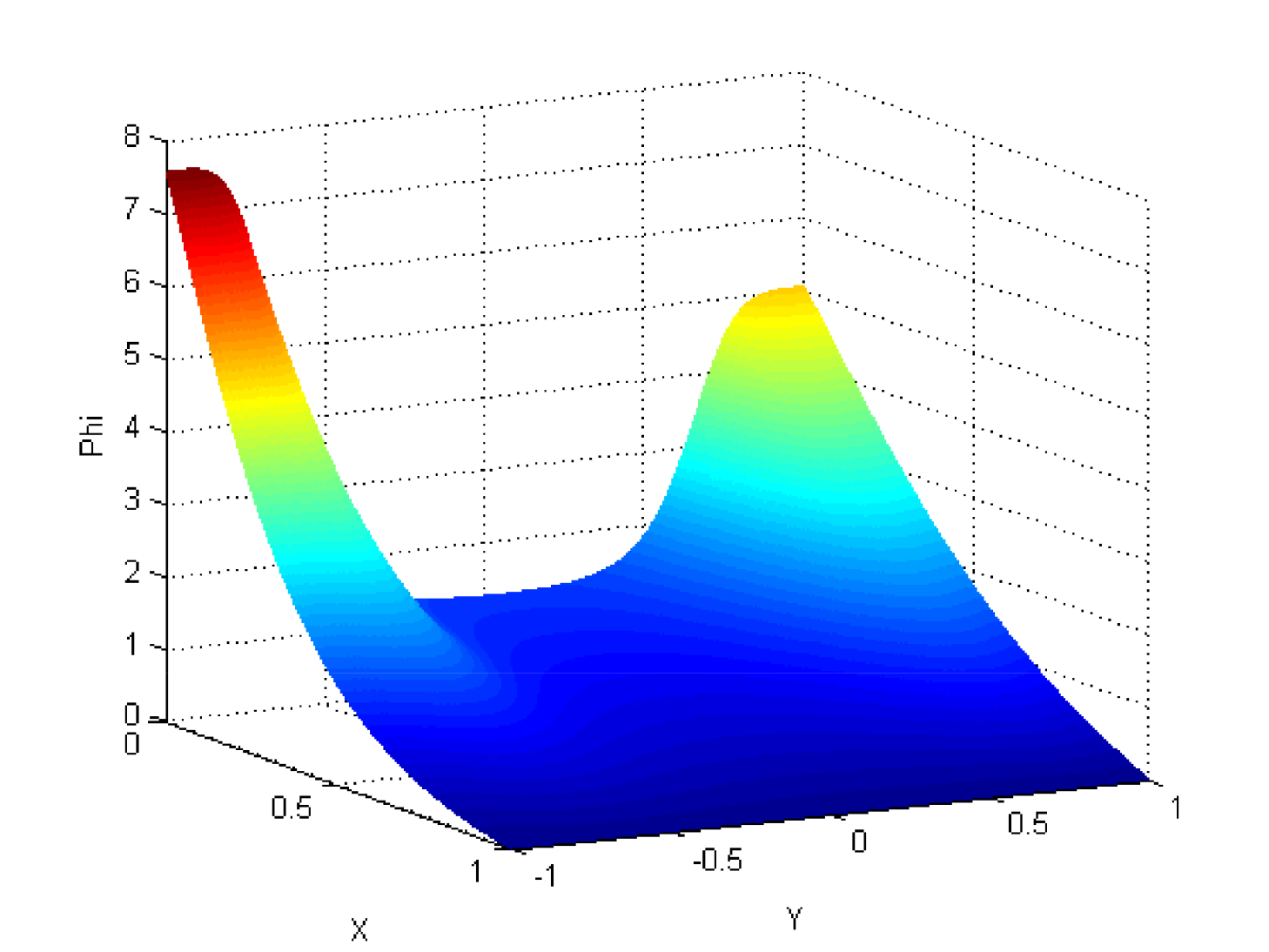}}
\hspace{0.2cm}
 \subfigure{
  \includegraphics[width=0.47\textwidth,height=0.28\textheight]{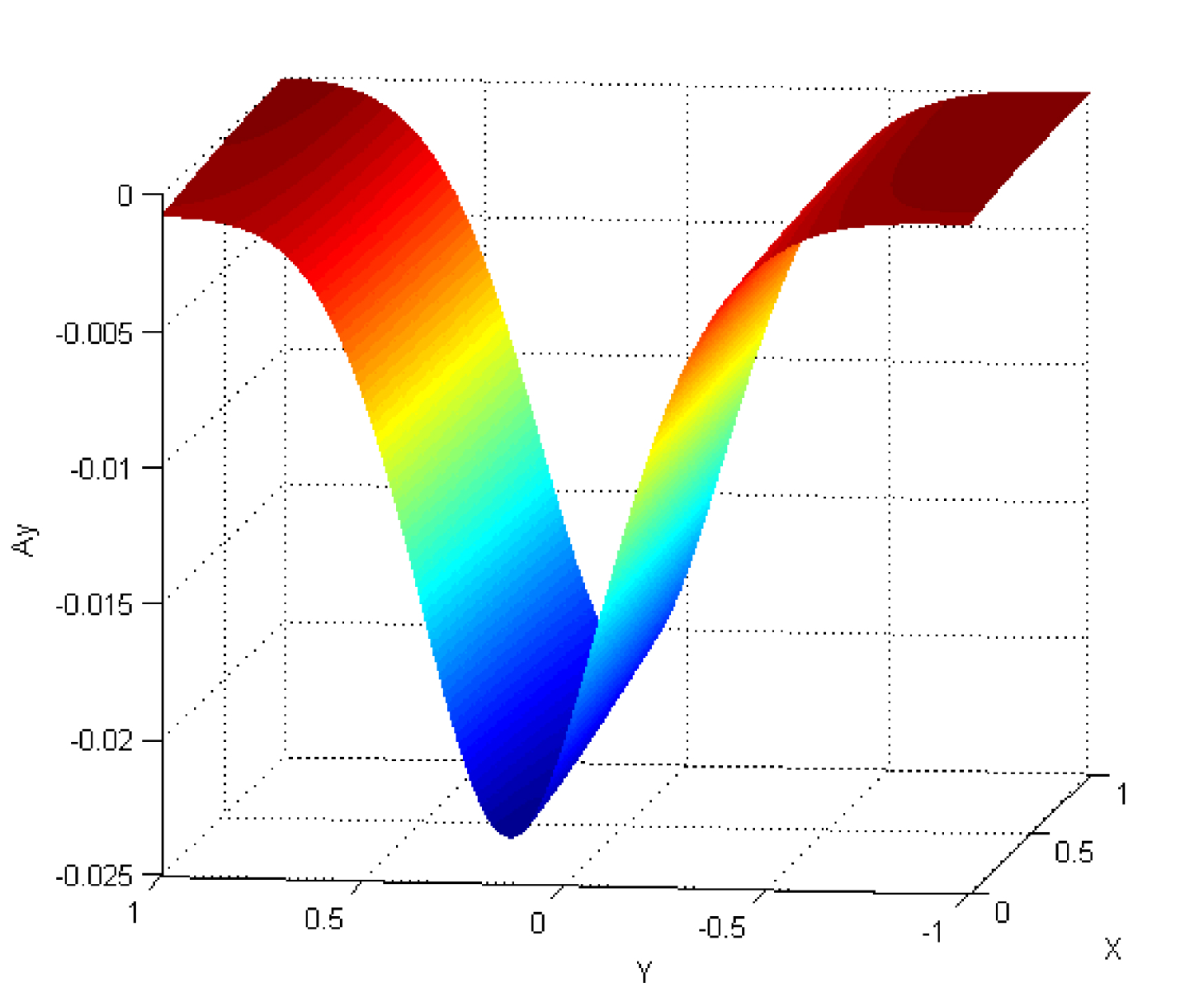}}
\end{center}  \vskip -5mm
  \caption{We represent the components $\phi$ and $A_{y}$ of the Yang$-$Mills field. \emph{Left} The profile of $\phi$. \emph{Right} The figure of $A_{y}$. The parameters are $m^{2}=-33/16$, $\mu=6.3$, $\epsilon=0.0$, $\sigma=0.5$, $\ell=3$, and $\beta=0.6$. We can see that these two figures are not symmetrical.}
\label{fig_scalar_warpfactor3}
\end{figure}

In this subsection, let us to continue to study the s+p-N-p Josephson junction. We tune the chemical potential $\mu(y)$ such that $\mu(-\infty)$ is in the s+p coexisting region $\mu_{c1}\sim\mu_{c2}$ and $\mu(+\infty)<\mu_{c1}$ is in the pure p-wave phase region. We still calculate  the phase difference $\gamma$  from Eq. (\ref{gamma1}).
The profiles of $A^{1}_{t}$ and $A_{y}^{1}$ are shown in Fig. \ref{fig_scalar_warpfactor3}, with the parameters $m^{2}=-33/16$, $\mu=6.3$, $\epsilon=0.0$, $\sigma=0.5$, $\ell=3$, and $\beta=0.6$.

\begin{figure}
  \begin{center}
  \subfigure{
  \includegraphics[width=0.47\textwidth,height=0.28\textheight]{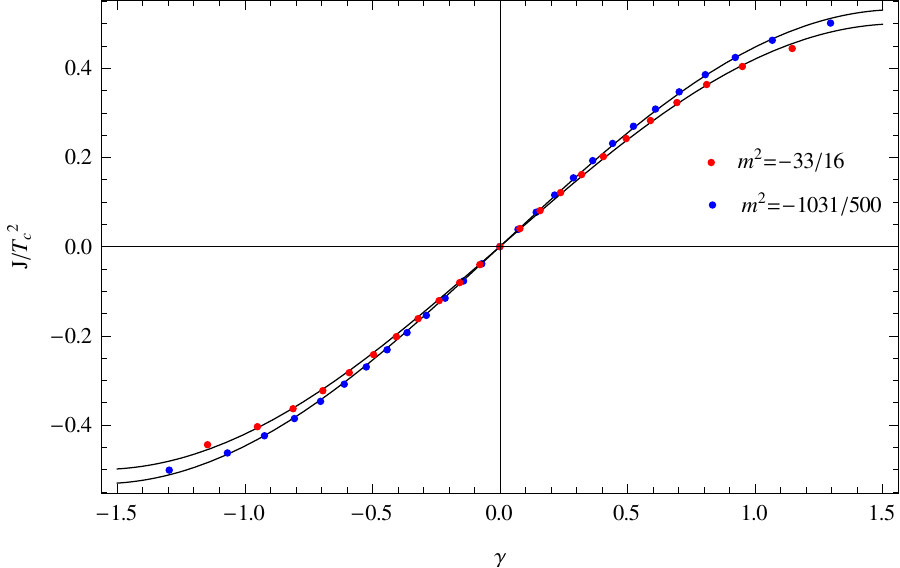}}
\hspace{0.2cm}
 \subfigure{
  \includegraphics[width=0.47\textwidth,height=0.28\textheight]{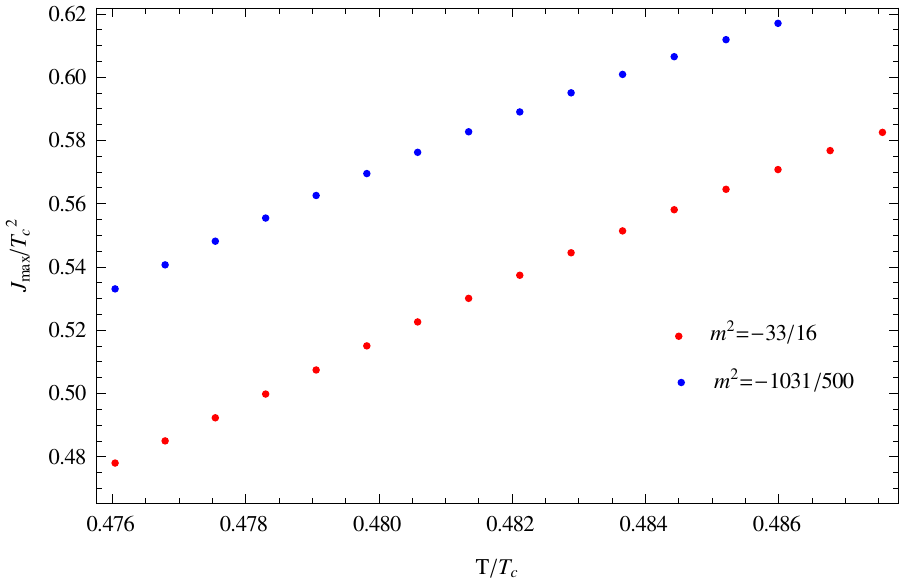}}
\end{center}  \vskip -5mm
  \caption{\emph{Left} The current $J/T_{c}^{2}$ as a sine function of the difference phase $\gamma$. \emph{Right} The $J_{max}/T_{c}^{2}$ increases with growing $T/T_{c}$. The black lines are sine curves and the points with $m^{2}=-33/16$(\emph{red}), $-1031/500$(\emph{blue}) are numerical results. The parameters are $\mu=6.3$, $\epsilon=0.0$, $\sigma=0.5$, $\ell=3$, and $\beta=0.6$.}
\label{fig_jgammaspnp jmaxtspnp}
\end{figure}

The dots are determined by the DC current $J$ and the phase difference $\gamma$ across the junction is shown on the left panel of Fig. \ref{fig_jgammaspnp jmaxtspnp}. With data fitting, we can see that $J$ is proportional to the sine of $\gamma$. We have
\begin{displaymath}
J/T_{c}^{2}\approx\left\{ \begin{array}{ll}
 0.4999sin\gamma & \qquad\textrm{for \quad$m^{2}=-33/16$},\\
 0.5318sin\gamma & \qquad\textrm{for \quad$m^{2}=-1031/500$}.
 \end{array} \right.
\end{displaymath}
From the figure, it is shown that when the value of $m^{2}$ increases the maximum current $J_{max}$ will grow. The dependence of $J_{max}$ on the temperature $T$ is shown on the right panel of Fig. \ref{fig_jgammaspnp jmaxtspnp}. The graph shows that $J_{max}/T_{c}^{2}$ increases with growing $T/T_{c}$, which is for the same reason as the case of the s+p-N-s+p junction.

\begin{figure}
  \begin{center}
  \subfigure{
  \includegraphics[width=0.47\textwidth,height=0.28\textheight]{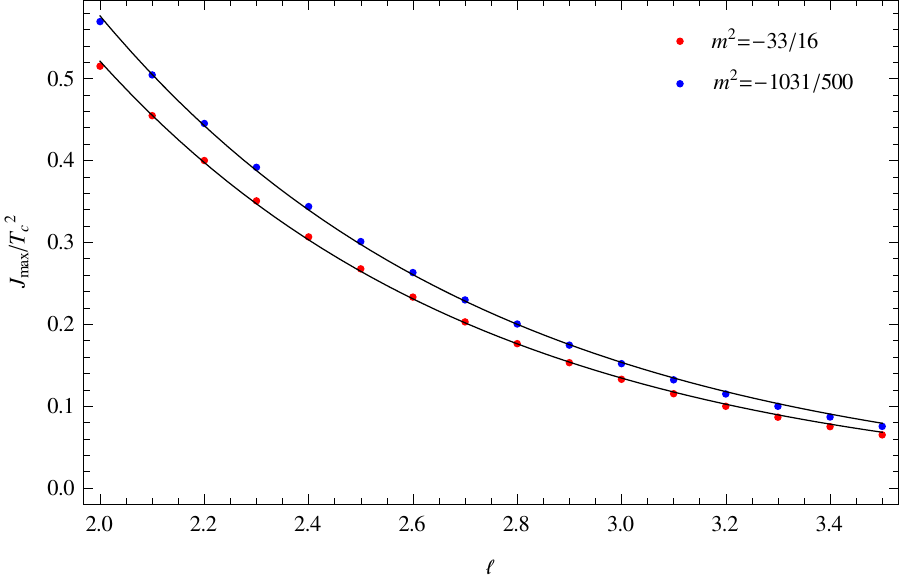}}
\hspace{0.2cm}
 \subfigure{
  \includegraphics[width=0.47\textwidth,height=0.28\textheight]{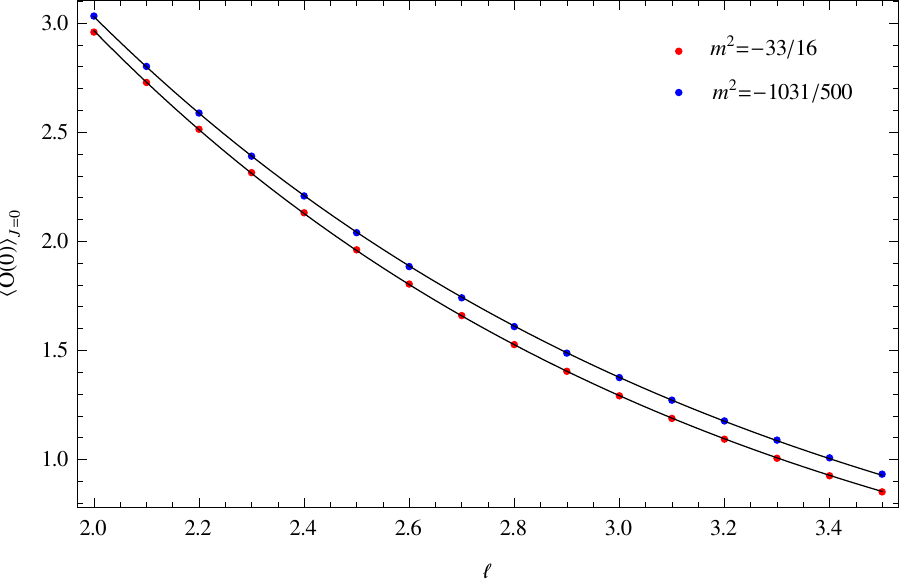}}
\end{center}  \vskip -5mm
  \caption{\emph{Left} The maximum current $J_{max}$ as an exponential function of $\ell$. \emph{Right} The total condensate $\langle\mathcal{O}(0)\rangle_{J=0}$ as an exponential function of $\ell$. Our numerical results are the points with $m^{2}=-33/16$(\emph{red}), $-1031/500$(\emph{blue}). In all the plots, we use $\mu=6.3$, $\epsilon=0.0$, $\sigma=0.5$, and $\beta=0.6$. The numerical data fit the exponential curves well.}
\label{fig_jamxlspnp olspnp}
\end{figure}

In the left panel of Fig. \ref{fig_jamxlspnsp olspnsp}, the dependence of   $J_{max}/T_{c}^{2}$ on $\ell$ is shown. We can see that The $J_{max}/T_{c}^{2}$   decays with $\ell$ exponentially and the change will be larger when $m^{2}$ becomes larger. Furthermore, we show the dependence of  $\langle\mathcal{O}(0)\rangle_{J=0}$ on $\ell$ in the right panel in Fig. \ref{fig_jamxlspnp olspnp}. $\langle\mathcal{O}(0)\rangle_{J=0}$ also decays with growing $\ell$ exponentially and $\langle\mathcal{O}(0)\rangle_{J=0}$ becomes larger with increasing $m^{2}$. We have
\begin{displaymath}
\left\{ \begin{array}{ll}
 J_{max}/T_{c}^{2}\approx7.843e^{-\ell/0.7376} \\
 \qquad\qquad\qquad\qquad  & \qquad\textrm{for \quad$m^{2}=-33/16$}.\\
 \langle\mathcal{O}(0)\rangle_{J=0}\approx15.61e^{-\ell/(2\times0.6021)}
 \end{array} \right.
\end{displaymath}
From the above result, we can obtain the coherence length (0.7376, 0.6021) from the above two equations, respectively. The error of the two values is about 18.4 \%. We have
\begin{displaymath}
\left\{ \begin{array}{ll}
 J_{max}/T_{c}^{2}\approx8.129e^{-\ell/0.7558} \\
 \qquad\qquad\qquad\qquad  & \qquad\textrm{for \quad$m^{2}=-1031/500$}.\\
 \langle\mathcal{O}(0)\rangle_{J=0}\approx14.68e^{-\ell/(2\times0.6337)}
 \end{array} \right.
\end{displaymath}
Similarly, the coherence length (0.7558, 0.6337) is obtained from the above two equations, respectively. The error of the two values is about 16.2 \%.

\subsection{s-N-p Josephson junction}
\begin{figure}
  \begin{center}
  \subfigure{
  \includegraphics[width=0.47\textwidth,height=0.28\textheight]{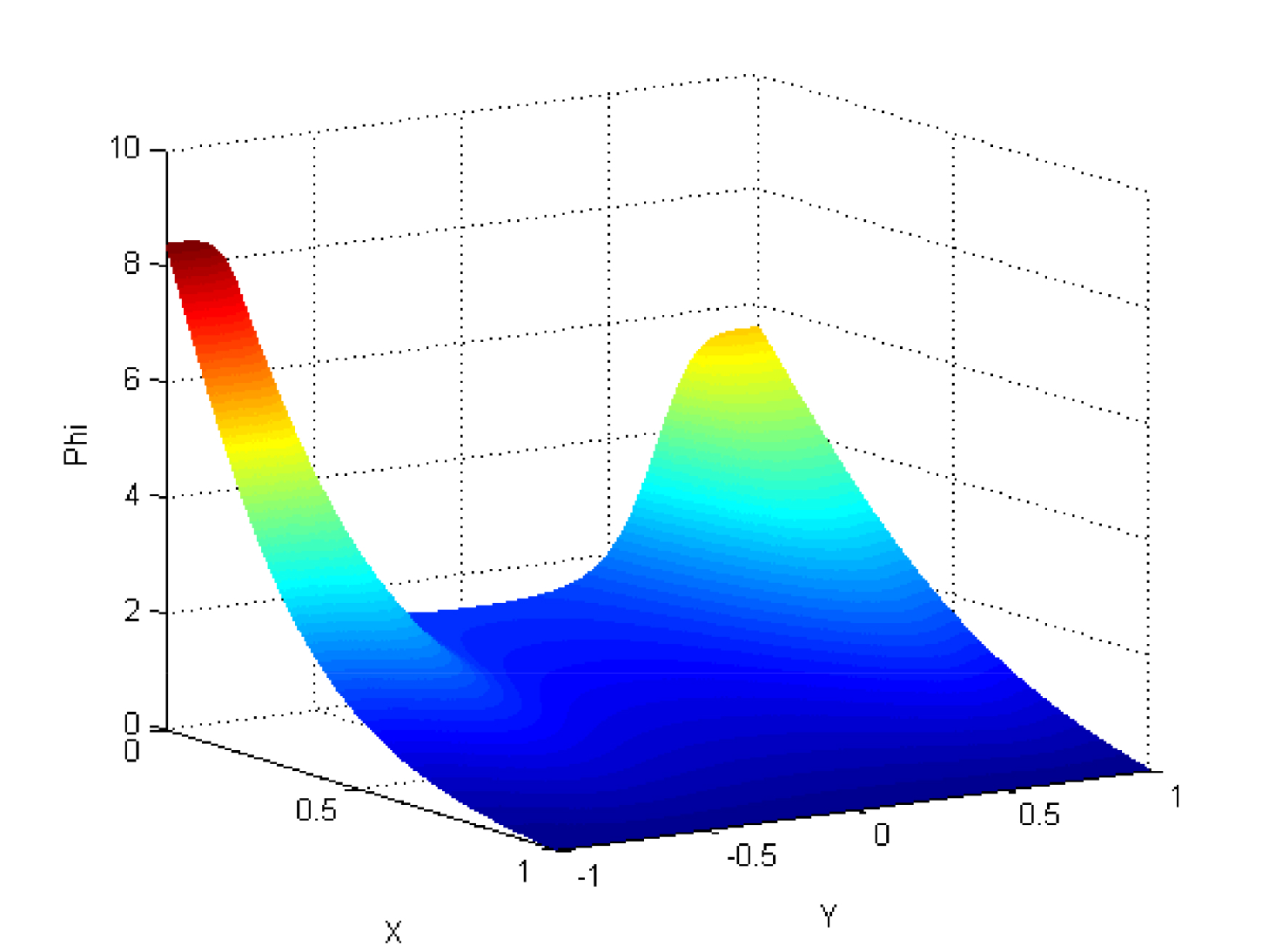}}
\hspace{0.2cm}
 \subfigure{
  \includegraphics[width=0.47\textwidth,height=0.28\textheight]{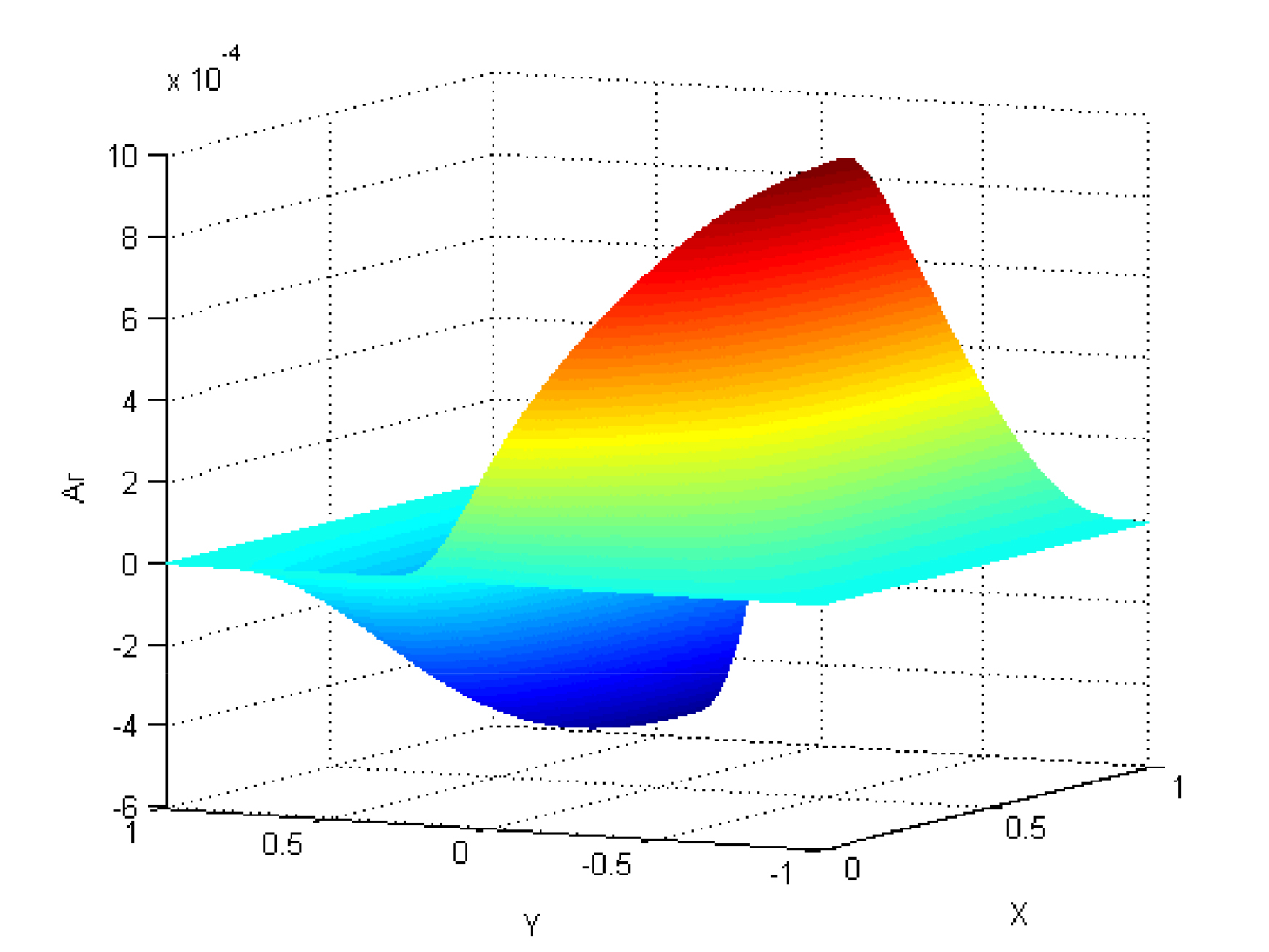}}
\end{center}  \vskip -5mm
  \caption{We represent the components $\phi$ and $A_{y}$ of the Yang$-$Mills field. \emph{Left} The profile of $\phi$. \emph{Right} The figure of $A_{y}$. The parameters are $m^{2}=-33/16$, $\mu=7$, $\epsilon=0.0$, $\sigma=0.5$, $\ell=3$, and $\beta=0.6$. We can see that these two figures are not symmetrical.}
\label{fig_scalar_warpfactor4}
\end{figure}

In this subsection, we will construct the  hybrid model of the s-wave and p-wave Josephson junction, namely, the s-N-p Josephson junction. We tune the value of the chemical potential $\mu(y)$ such that $\mu(-\infty)>\mu_{c2}$ is in the pure s-wave region  and $\mu(+\infty)<\mu_{c1}$ is in the pure p-wave phase region. The phase difference $\gamma$ can be obtained in Eq. (\ref{gamma1}). The profiles of $A^{1}_{t}$ and $A_{y}^{1}$ are shown in Fig. \ref{fig_scalar_warpfactor4}, with the parameters $m^{2}=-33/16$, $\mu=7$, $\epsilon=0.0$, $\sigma=0.5$, $\ell=3$, and $\beta=0.6$.

\begin{figure}
  \begin{center}
  \subfigure{
  \includegraphics[width=0.47\textwidth,height=0.28\textheight]{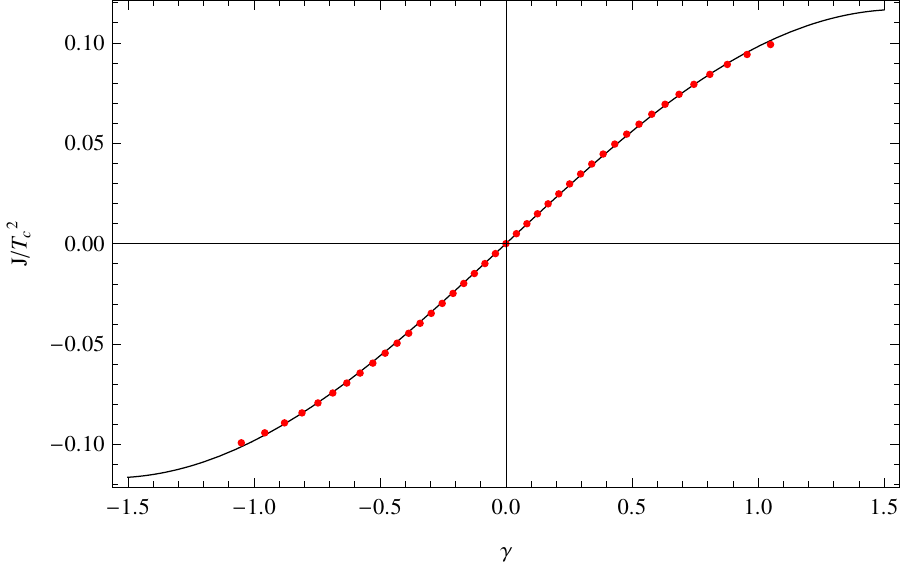}}
\hspace{0.2cm}
 \subfigure{
  \includegraphics[width=0.47\textwidth,height=0.28\textheight]{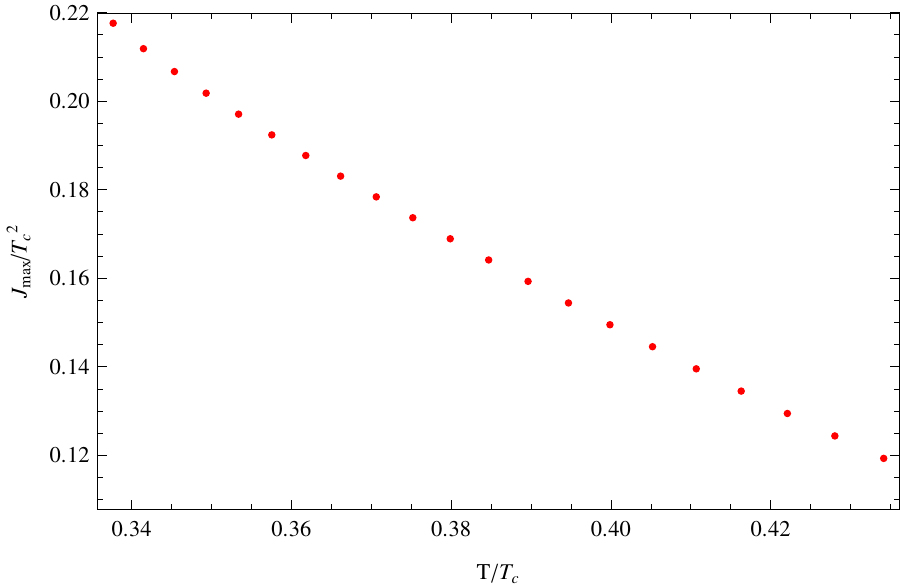}}
\end{center}  \vskip -5mm
  \caption{\emph{Left} The current $J/T_{c}^{2}$ as a sine function of $\gamma$. \emph{Right} $J_{max}/T_{c}^{2}$ decays with growing $T/T_{c}$. In all the plots, we use $m^{2}=-33/16$, $\mu=7$, $\epsilon=0.0$, $\sigma=0.5$, $\ell=3$, and $\beta=0.6$. The numerical data fit the exponential curves well.}
\label{fig_jgammasnp jmaxtsnp}
\end{figure}

We show the dependence of the current $J$ on the phase difference $\gamma$ across the junction  on the left side of Fig. \ref{fig_jgammasnp jmaxtsnp}, in which it is shown that $J$ is proportional to the sine of $\gamma$. We have

\be
J/T_{c}^{2}\approx0.1166sin\gamma,\qquad for\quad m^{2}=-33/16.
\ee
The dependence of $J_{max}$ on the temperature $T$ is shown on the right panel of Fig. \ref{fig_jgammasnp jmaxtsnp}, which shows that $J_{max}/T_{c}^{2}$ decreases with growing $T/T_{c}$.

\begin{figure}
  \begin{center}
  \subfigure{
  \includegraphics[width=0.47\textwidth,height=0.28\textheight]{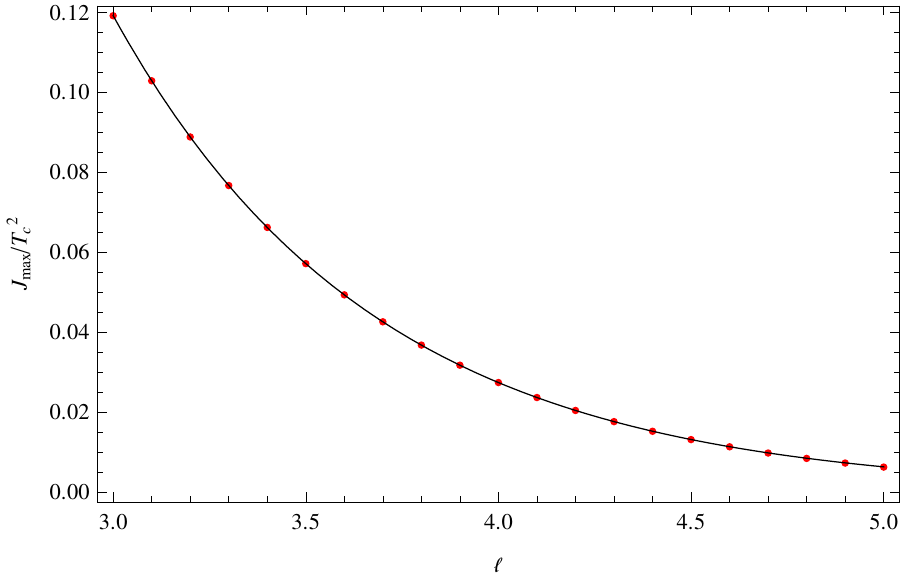}}
\hspace{0.2cm}
 \subfigure{
  \includegraphics[width=0.47\textwidth,height=0.28\textheight]{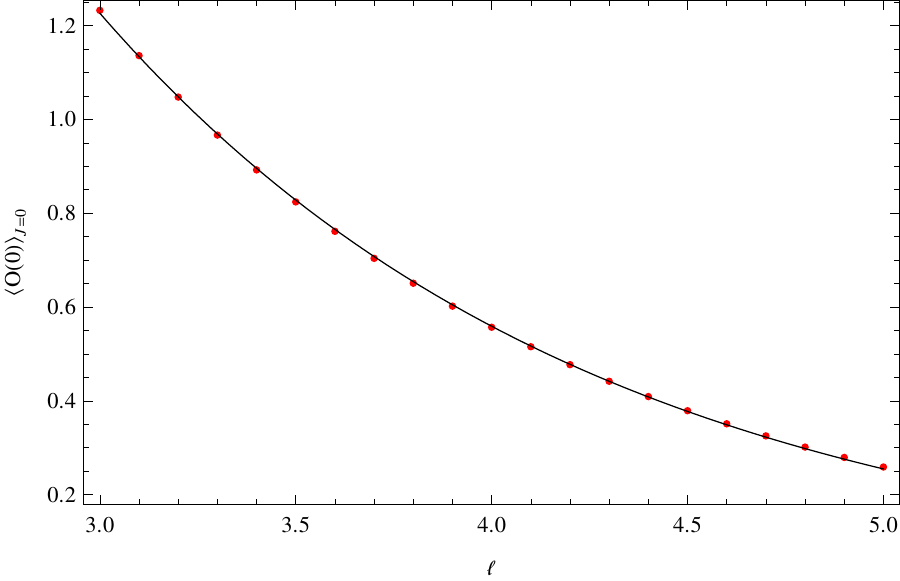}}
\end{center}  \vskip -5mm
  \caption{\emph{Left} The maximum current $J_{max}/T_{c}^{2}$ as an exponential function of $\ell$. \emph{Right} The total condensate $\langle\mathcal{O}(0)\rangle_{J=0}$ as exponential function of $\ell$. In all the plots, we use $m^{2}=-33/16$, $\mu=7$, $\epsilon=0.0$, $\sigma=0.5$, and $\beta=0.6$. The numerical data fit the exponential curves well.}
\label{fig_jamxlsnp olsnp}
\end{figure}

The dependence of $J_{max}$  on the width $\ell$ of the gap is shown on the left panel of Fig. \ref{fig_jamxlsnp olsnp}, which  shows that $J_{max}/T_{c}^{2}$ decays with growing $\ell$ exponentially. The dependence of $\langle\mathcal{O}(0)\rangle_{J=0}$ on the width $\ell$ of the gap is shown on the right panel of Fig. \ref{fig_jamxlsnp olsnp}, which  predicts that $\langle\mathcal{O}(0)\rangle_{J=0}$ decays with growing $\ell$ exponentially. We have
\begin{displaymath}
\left\{ \begin{array}{ll}
 J_{max}/T_{c}^{2}\approx9.842e^{-\ell/0.6798} \\
 \qquad\qquad\qquad\qquad  & \qquad\textrm{for \quad$m^{2}=-33/16$}.\\
 \langle\mathcal{O}(0)\rangle_{J=0}\approx12.95e^{-\ell/(2\times0.6362)}
 \end{array} \right.
\end{displaymath}
The coherence length (0.6798, 0.6362) can be obtained from these equations. The error of these two values is about 6.4 \%.\\

\section{Conclusion and discussion}
In this paper, we set up a holographic model for a hybrid s-wave and p-wave DC Josephson junction with a scalar triplet charged under the $SU(2)$ gauge field in the background of a (\(3+1\))-dimensional AdS black hole. We get a set of partial differential equations of fields that are nonlinear and coupled and solve them numerically. We construct a new chemical potential $\mu(y)$ and tune the parameters in it, so the s+p-N-s+p junction, s+p-N-s junction, s+p-N-p junction, and s-N-p junction can be obtained, respectively. For the four kinds of junctions, we find that the DC is proportional to the sine of the phase difference across the junction and the coherence lengths are different. We also study the relationship between $J_{max}/T_{c}^{2}$ and $\ell$, the total condensation $\langle\mathcal{O}(0)\rangle_{J=0}$ and $\ell$, $J_{max}/T_{c}^{2}$ and $T/T_{c}$, respectively.

The reason we take $m^{2}=-33/16$ is that when $m^{2}<-33/16$, the region of s+p coexistence is too small, when $m^{2}>-33/16$, the value in the region of s+p coexistence is too large for the junction, the numerical results are not good. It is well known that the Josephson period is $2\pi$ in the $p_{y}$ wave s-N-p junction, so our ansatz just corresponds to the $p_{y}$ wave junction. To our surprise, the periods of the current are also $2\pi$ in the remaining three kinds of junctions. For the s+p-N-s+p junction and the s+p-N-p junction, we take different values of $m^{2}=-33/16, -1031/500$ and find that when the value becomes larger, the $J$, $J_{max}/T_{c}^{2}$ and $\langle\mathcal{O}(0)\rangle_{J=0}$ will become larger. The reason we take another value of $m^{2}=-1031/500$ is that the region of s+p coexistence is small, we should take the other value of $m^{2}$ as it approaches the $m^{2}=-33/16$. The maximum current increases with the growing temperature in the s+p-N-s+p and the s+p-N-p junction. Except the s+p-N-s+p junction, the phase difference should be obtained by $\gamma=-\int_{-\infty}^{c}dy[\nu(y)-\nu(-\infty)]-\int_{c}^{+\infty}dy[\nu(y)-\nu(+\infty)]$ without loss of generality. When $c\neq0$, the relationship between the current and the phase difference is $J/T_{c}^{2}=(J_{max}/T_{c}^{2})sin(\gamma+\phi)$, where $\phi$ is the origin's phase difference and $\phi\neq0$. So it is more convenient to set $c=0$ to make $\phi=0$.

Note that our model also can describe the s-wave superconductor, the p-wave superconductor, the s+p coexistence superconductor, the s-wave junction, and the p-wave junction. The p-wave contains a $p_{x}$ wave and a $p_{y}$ wave, the Josephson periods are $\pi$ and $2\pi$, respectively. In the present study, this ansatz just describes the $p_{y}$ wave. So, it should be of great interest to construct an ansatz which can describe a $p_{x}$ wave and a $p_{y}$ wave, respectively.

Finally, we have studied the hybrid and coexisting s-wave and p-wave junctions with the probe limit, and we would like to  study these kinds of junctions with the gravity backreaction in the future. So far, we have studied the hybrid and coexisting s-wave and p-wave DC Josephsons junction in the AdS black hole background. It is intended to study these DC  junctions by taking an AdS soliton as the geometry background in our further work.

\section*{Acknowledgement}
YQW would like to thank Li Li and Hai-Qing Zhang for very helpful discussion. SL and YQW were supported by the National Natural Science
Foundation of China.

\end{document}